# THE HELMHOLTZ THEOREM AND SUPERLUMINAL SIGNALS

V.P. Oleinik

Department of General and Theoretical Physics,
National Technical University of Ukraine "Kiev Polytechnic Institute",
Prospect Pobedy 37, Kiev, 03056, Ukraine;
e-mail: valoleinik@users.ntu-kpi.kiev.ua; yuri@arepjev.relc.com
http://www.chronos.msu.ru/lab-kaf/Oleynik/eoleynik.html

The conventional decomposition of a vector field into longitudinal (potential) and transverse (vortex) components (Helmholtz's theorem) is claimed in [1] to be inapplicable to the time-dependent vector fields and, in particular, to the retarded solutions of Maxwell's equations. Because of this, according to [1], a number of conclusions drawn in [2] on the basis of the Helmholtz theorem turns out to be erroneous. The Helmholtz theorem is proved in this letter to hold for arbitrary vector field, both static and time-dependent. Therefore, the conclusions of the paper [2] questioned in [1] are true. The validity of Helmholtz's theorem in the general case is due to the fact that the decomposition above of vector field does not influence the field time coordinate, which plays, thus, a passive role in the decomposition procedure. An analysis is given of the mistakes made in [1]. It is noted that for point particle the longitudinal and transverse components of electric field, taken separately, are characterized by the infinitely great velocity of propagation. However, superluminal contributions to the expression for the total electric field cancel each other.

## Decomposition of a vector field into longitudinal and transverse components (general considerations)

As is known [3], the simplest way of separating a vector field

$$A(x) = A(\mathbf{r},t) \tag{1}$$

into longitudinal and transverse components is based on the Fourier expansion of the function $A(\mathbf{r},t)$:

$$A(\mathbf{r},t) = \int d^3 p \, \exp(i\mathbf{pr}) \, \widetilde{A}(\mathbf{p},t). \tag{2}$$

Let us represent the function $\widetilde{A}(\mathbf{p},t) \equiv \widetilde{A}$ as a sum

$$\widetilde{A} = \widetilde{A}_{\parallel} + \widetilde{A}_{\perp}, \tag{3}$$

where

$$\widetilde{A}_{\parallel} \parallel \mathbf{p} \quad \text{and} \quad \widetilde{A}_{\perp} \perp \mathbf{p}. \tag{4}$$

As is seen from (3) and (4),

$$(\mathbf{p}\widetilde{A}) = (\mathbf{p}\widetilde{A}_{\parallel}), \quad [\mathbf{p}\widetilde{A}] = [\mathbf{p}\widetilde{A}_{\perp}]. \tag{5}$$

Putting in accordance with (4) $\widetilde{A}_{\parallel} = c\mathbf{p}$, where $c = c(\mathbf{p})$ is a scalar, with the help of the first of equalities (5) one can derive:

$$c = \frac{(\mathbf{p}\widetilde{A})}{\mathbf{p}^2}, \quad \widetilde{A}_{\parallel}(\mathbf{p},t) = \frac{\mathbf{p}(\mathbf{p}\widetilde{A})}{\mathbf{p}^2}. \tag{6}$$

With the help of the second of equalities (5) one can get:

$$[\mathbf{p}[\mathbf{p}\widetilde{A}]] = -\mathbf{p}^2 \widetilde{A}_{\perp}, \quad \widetilde{A}_{\perp}(\mathbf{p},t) = -\frac{[\mathbf{p}[\mathbf{p}\widetilde{A}]]}{\mathbf{p}^2}. \tag{7}$$

From the expressions (2), (3), (6) and (7) follow the equalities ($A(\mathbf{r},t) = A_{\parallel}(\mathbf{r},t) + A_{\perp}(\mathbf{r},t)$):

$$A_{\|}(r,t) = \int d^3p \exp(ipr) \, \widetilde{A}_{\|}(p,t) = -grad \, div \int d^3p \exp(ipr) \frac{\widetilde{A}(p,t)}{p^2}, \quad (8)$$

$$A_{\perp}(r,t) = \int d^3p \exp(ipr) \, \widetilde{A}_{\perp}(p,t) = curl \, curl \int d^3p \exp(ipr) \frac{\widetilde{A}(p,t)}{p^2}. \quad (9)$$

Using formula (2) and Fourier expansion

$$\frac{1}{|r-r'|} = \frac{1}{2\pi^2} \int d^3p \frac{1}{p^2} \exp(ip(r-r')),$$

one can arrive at the following relationship:

$$\int d^3p \exp(ipr) \frac{\widetilde{A}(p,t)}{p^2} = \frac{1}{4\pi} \int d^3x' \frac{1}{|r-r'|} A(r',t). \quad (10)$$

From (8) and (10) we obtain the known formula:

$$A_{\|}(r,t) = -grad \, div \frac{1}{4\pi} \int d^3x' \frac{1}{|r-r'|} A(r',t). \quad (11)$$

Analogous formula can be obtained for transverse component as well.

Note that if we decompose the function $\widetilde{A}(p,t)$ in (2) into Fourier integral in time coordinate,

$$\widetilde{A}(p,t) = \int dp_0 \exp(-ip_0 t) \, \widetilde{\widetilde{A}}(p,p_0), \quad (12)$$

then the Fourier expansions (2), (8), and (9) can be written as (see (6) and (7)):

$$A(x) = A_{\|}(x) + A_{\perp}(x) = \int d^4p \exp(-ipx) \, \widetilde{\widetilde{A}}(p), \quad \widetilde{\widetilde{A}}(p) = \widetilde{\widetilde{A}}(p,p_0), \quad px = p_0 t - pr, (13)$$

$$A_{\|}(x) = \int d^4p \exp(-ipx) \, \widetilde{\widetilde{A}}_{\|}(p), \quad \widetilde{\widetilde{A}}_{\|}(p) = \widetilde{\widetilde{A}}_{\|}(p,p_0) = \frac{p(p\widetilde{\widetilde{A}}(p,p_0))}{p^2}, \quad (14)$$

$$A_{\perp}(x) = \int d^4p \exp(-ipx) \, \widetilde{\widetilde{A}}_{\perp}(p), \quad \widetilde{\widetilde{A}}_{\perp}(p) = \widetilde{\widetilde{A}}_{\perp}(p,p_0) = -\frac{[p[p\widetilde{\widetilde{A}}(p,p_0)]]}{p^2}. \quad (15)$$

As is seen from the general consideration given above, the formulas obtained for longitudinal and transverse components hold for any arbitrary function $A(x)$ (1). The function $A(x)$ can correspond, in particular, to both the retarded and advanced solutions of Maxwell's equations and to any superposition of these solutions as well. The only requirement imposed on the function $A(x)$ is that it be decomposable into Fourier integral (see (2) and (13)). It is claimed in [1], without any grounds, that for retarded solutions of Maxwell's equations the decomposition of the kind (13) is incorrect. This assertion is erroneous, as it is evident from the general theory given above.

Before proceeding to an analysis of mistakes in [1], let us prove the following lemma.

**Lemma.** The longitudinal component of the vector field $A(x)$, defined by

$$A(x) = \int d^4x' f(x-x') J(x'), \quad (16)$$

where $f(x-x')$ and $J(x)$ are arbitrary functions decomposable into Fourier integrals,

$$f(x-x') = \int d^4k \exp(-ik(x-x')) \widetilde{f}(k), \quad (17)$$

$$J(x) = \int d^4k \exp(-ikx) \widetilde{J}(k), \quad (18)$$

is given by

$$A_{\|}(x) = \int d^4x' f(x-x') J_{\|}(x'), \quad (19)$$

with $J_{\|}(x)$ being the longitudinal component of vector $J(x)$ defined in the usual fashion, that is defined according to one of the formulas (8), (11), (14).

**Proof.** Expanding the function $A(x)$ (16) into Fourier integral (13), we calculate its Fourier transform:

$$\widetilde{\widetilde{A}}(p) = \frac{1}{(2\pi)^4} \int d^4x \exp(ipx) A(x) = \frac{1}{(2\pi)^4} \int d^4x \exp(ipx) \int d^4x' f(x-x') J(x').$$

Substituting Fourier expansions (17) and (18) into the last formula and making elementary transformations, we get:

$$\widetilde{\widetilde{A}}(p) = (2\pi)^4 \widetilde{f}(p) \widetilde{J}(p). \tag{20}$$

Hence it follows, in view of (14), that

$$\widetilde{\widetilde{A}}_{||}(p) = (2\pi)^4 \widetilde{f}(p) \widetilde{J}_{||}(p). \tag{21}$$

Making use of (21) and calculating the Fourier integral according to (14), we arrive at the sought-for expression (19).

**Note.** The lemma is seen from the proof presented to hold for arbitrary function $f(x-x')$ decomposable into Fourier integral. Obviously, both the retarded and advanced Green functions and also any their linear combination can be used as the function $f(x-x')$ in (19). The Helmholtz theorem is claimed in [1] to be inapplicable to a time-dependent vector field. That the statement is erroneous is evident from the general considerations given above: according to equalities (8) and (9), the time coordinate of function $A(x)$ plays merely a role of a parameter which does not influence the mathematical procedure of decomposing the vector field into longitudinal and transverse components.

## Analysis of mistakes and discussion

In order to find a mistake in [1], let us consider the retarded solution for electric field $E$, which is written in [1] as ($D_R$ is the retarded Green function):

$$E = E_{||} + E_{\perp} + E_{\tau}, \tag{22}$$

where

$$E_{||}(x) = -4\pi \vec{\nabla} \int d^4x' D_R(x-x') \rho(x'), \tag{23}$$

$$E_{\perp}(x) = -\vec{\nabla} \times \int d^4x' D_R(x-x') \frac{\partial B(x')}{\partial t'}, \tag{24}$$

$$E_{\tau}(x) = \vec{\nabla} \times \int d^4x' D_R(x-x') \frac{\partial B(x')}{\partial t'} -$$
$$- 4\pi \frac{\partial}{\partial t} \int d^4x' D_R(x-x') J(x'). \tag{25}$$

Note that formulas (23), (24), and (25) in the above are numbered in [1] correspondingly as (8), (9), and (10). Quantities $E_{||}$, $E_{\perp}$, and $E_{\tau}$ are termed in [1] correspondingly longitudinal, transverse, and temporal components. These terms cannot provoke any objections. However, it should be borne in mind that the field $E_{||}$ (23) is merely a part of longitudinal component of the total field $E$, since longitudinal component is contained also in the field $E_{\tau}$ (25). Really, according to Lemma (see formula (19))**,**

$$E_{\tau||}(x) = -4\pi \frac{\partial}{\partial t} \int d^4x' D_R(x-x') J_{||}(x'). \tag{26}$$

Hence, the total longitudinal component equals the sum of (23) and (26). So, **the true expression for longitudinal component is given**, as it should be, **by Rohrlich's formula** (see formula (2) in [1]):

$$E_{||}^R(x) = -4\pi \int d^4x' D_R(x-x') \left[ \frac{\partial J_{||}(x')}{\partial t'} + \vec{\nabla}' \rho(x') \right]. \tag{27}$$

Separating, then, the potential (longitudinal) component from Maxwell's equation

$$[\vec{\nabla} B] = \frac{\partial E}{\partial t} + 4\pi J,$$

we obtain equation

$$4\pi J_{\|} = -\frac{\partial E_{\|}}{\partial t}. \tag{28}$$

With the help of the last equality and Gauss's theorem $\vec{\nabla} E = 4\pi\rho$, $\vec{\nabla} E = \vec{\nabla} E_{\|}$, we can derive:

$$\frac{\partial J_{\|}}{\partial t} + \vec{\nabla}\rho = -\frac{1}{4\pi}\left(\frac{\partial^2}{\partial t^2} - \vec{\nabla}^2\right) E_{\|}, \qquad \left([\vec{\nabla}[\vec{\nabla} E_{\|}]] = \vec{\nabla}(\vec{\nabla} E_{\|}) - \vec{\nabla}^2 E_{\|} = 0\right), \tag{29}$$

where, obviously,

$$E_{\|}(x) = -\vec{\nabla}\varphi(x), \quad \varphi(x) = \varphi(r,t) = \int d^3 x' \frac{\rho(r',t)}{|r-r'|}. \tag{30}$$

Substitution of (29) into (27) results after simple manipulations in Rohrlich's formula (see (4) in [1]):

$$E_{\|}^{R}(x) = -\vec{\nabla}\varphi(x). \tag{31}$$

As is seen from (30) and (31), longitudinal component of retarded electric field propagates at the infinitely great velocity. Strange as it may seem at the first sight, this conclusion follows strictly from Maxwell's equations [4,5]. Indeed, one can derive from Maxwell's equations the following wave equation:

$$\left(\frac{\partial^2}{\partial t^2} - \vec{\nabla}^2\right) E = -4\pi\left(\frac{\partial J}{\partial t} + \vec{\nabla}\rho\right). \tag{32}$$

Its longitudinal and transverse components are of the form:

$$\left(\frac{\partial^2}{\partial t^2} - \vec{\nabla}^2\right) E_{\|} = -4\pi\left(\frac{\partial J_{\|}}{\partial t} + \vec{\nabla}\rho\right). \tag{33}$$

$$\left(\frac{\partial^2}{\partial t^2} - \vec{\nabla}^2\right) E_{\perp} = -4\pi\frac{\partial J_{\perp}}{\partial t}. \tag{34}$$

Taking into account the equality

$$\frac{\partial^2 E_{\|}}{\partial t^2} = -4\pi\frac{\partial J_{\|}}{\partial t},$$

which results from (28), we can derive from (33) the wave equation

$$\vec{\nabla}^2 E_{\|} = 4\pi\vec{\nabla}\rho. \tag{35}$$

Since for point particle the right hand side of Eq. (35) is localized at the point where the particle is placed, then from (35) it follows that the longitudinal component of electric field propagates at the infinitely great velocity. This conclusion follows, too, from Eq. (33), although there stands in its left hand side the D'Alembert operator corresponding to the wave propagation at the light velocity in vacuum. This is seen from the fact that the function $J_{\|}$ in the right hand side of (33) is not localized at the point occupied by particle but is "smeared" over the whole space [6]. The transverse component of electric field also propagates at the infinitely great velocity (this is seen from Eq. (34), the right hand side of which contains transverse component of the current density $J_{\perp}$ distributed over the whole space). The last statement is not in conflict with the fact that transverse electromagnetic waves propagate at light velocity since electromagnetic field contains, besides transverse waves (the photon component of the field), the own field (the self-field) of electrically charged particles (the purely classical component of the field) [4-6]. It should be emphasized that for a point particle the total electric field $E$ propagates at the light velocity, as it follows from (32). This means that superluminal contributions to the components $E_\parallel$ and $E_\perp$ cancel each other in the expression for the total field $E$ [4,5].

Note that from the expression for total electric field

$$E(x) = -4\pi \int d^4x' D_R(x-x') \left[ \frac{\partial J(x')}{\partial t'} + \vec{\nabla}'\rho(x') \right]$$

(see Eq. (1) in [1]) one can easily derive the following representation for transverse component of electric field:

$$E_\perp(x) = -4\pi \int d^4x' D_R(x-x') \frac{\partial J_\perp(x')}{\partial t'}. \tag{36}$$

Combining (31) and (36), one can obtain the total electric field (cf. [1] and [2]):

$$E = -\vec{\nabla}\varphi + E_\perp.$$

In conclusion, it should be emphasized that the Helmholtz theorem holds for arbitrary physical solutions of Maxwell's equations and hence the conclusions of the paper [2] questioned in [1] are true. Contrary to the statements in [1], the expressions (23) and (24) cannot be taken as longitudinal and transverse components, respectively, of electric field because each of them is merely a part of corresponding component. Correct expressions for the components are given by (27) for the longitudinal component and by (36) for the transverse one.

The author is grateful to Prof. F. Rohrlich for discussing the problem and Dr. Yu.D.Arepjev for interest in the paper and many fruitful discussions.